\documentstyle[prl,aps,multicol,epsfig]{revtex}

\begin{document}

\draft

\title{Spin-orbit coupling in interacting
quasi-one-dimensional electron systems}

\author{A.~V.~Moroz, K.~V.~Samokhin$^*$, and C.~H.~W.~Barnes}
\address{Cavendish Laboratory,
University of Cambridge, \\ Madingley Road, Cambridge CB3 0HE,
United Kingdom}

\date{\today}
\maketitle

\begin{abstract}
We present a new model for the study of spin-orbit coupling in
interacting quasi-one-dimensional systems and solve it exactly to
find the spectral properties of such systems.  We show that the
combination of spin-orbit coupling and electron-electron
interactions results in: the replacement of separate spin and
charge excitations with two new kinds of bosonic mixed-spin-charge
excitation, and a characteristic modification of the spectral
function and single-particle density of states.  Our results show
how manipulation of the spin-orbit coupling, with external
electric fields, can be used for the experimental determination of
microscopic interaction parameters in quantum wires.
\end{abstract}

\begin{multicols}{2}
\narrowtext

In contemporary condensed matter physics there exists a great
variety of electron systems that can be considered as
quasi-one-dimensional (Q1D). Among them are conducting
polymers~\cite{polymers}, carbon nanotubes~\cite{tubes}, and
semiconductor heterostructures~\cite{sp}. They combine the
richness of observable physical properties with the possibility of
finding exact solutions to non-trivial interacting problems. The
major theoretical advance in this field was the formulation and
solution of the Tomonaga-Luttinger model~\cite{TL}, which revealed
features that are generic to many interacting Q1D electron
systems, such as separation of charge and spin degrees of freedom
and the anomalous scaling of correlation functions~\cite{Voit}.

\begin{figure}
\begin{center}
\epsfig{file=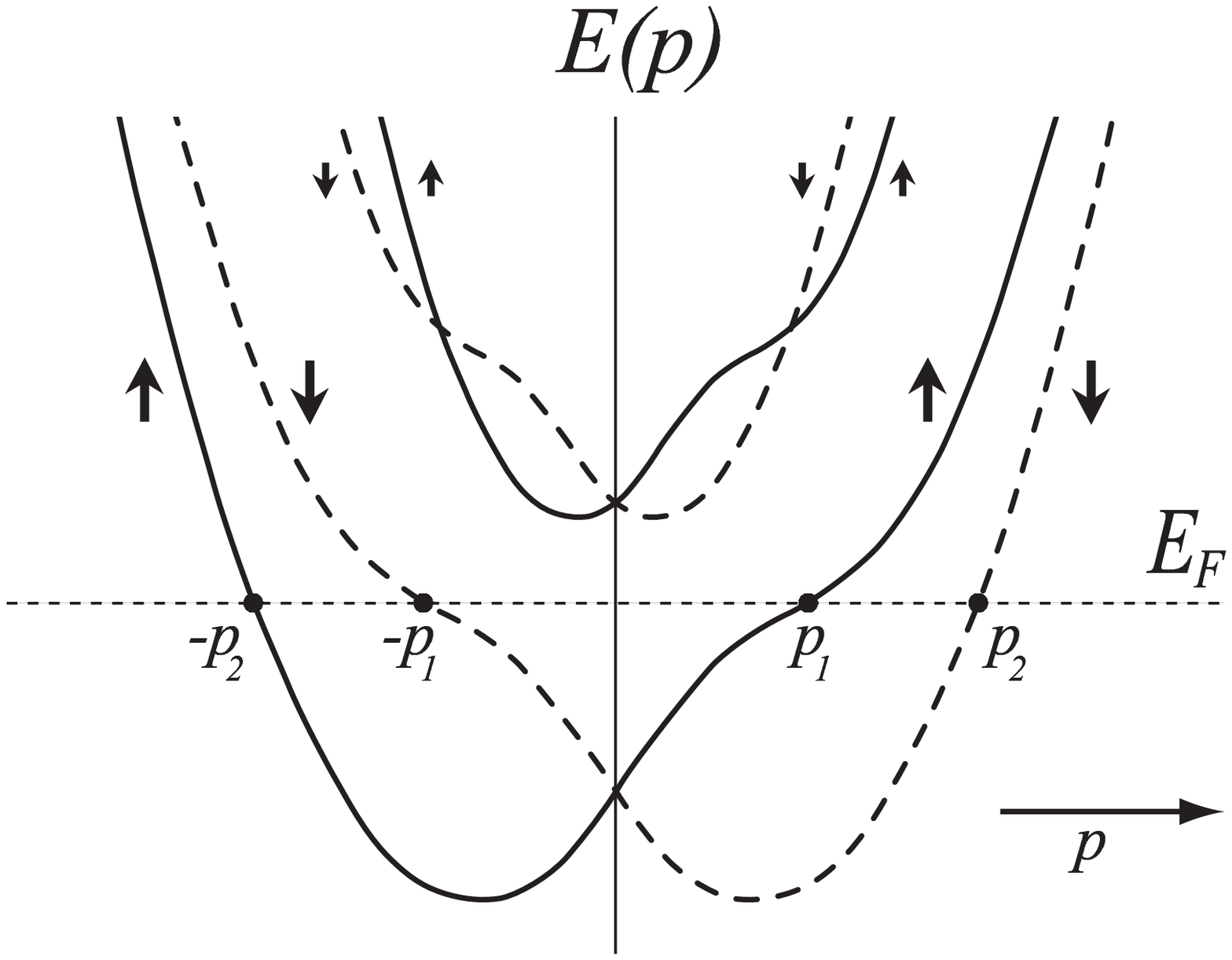,width=3.00in,height=2.00in}
\caption{Single-electron energy spectrum in a quantum wire with
spin-orbit interaction. The dotted line is the Fermi energy level
$E=E_F$ and $\pm p_{1,2}$ are the Fermi momenta of respective
groups of electrons.} \label{fig1}
\end{center}
\end{figure}

The very existence of Q1D systems (or quantum wires) arises
essentially from confining electrical forces, which prevent the
lateral motion of electrons. The accompanying electric fields
necessarily give rise to a coupling between the spin and orbital
degrees of freedom of propagating electrons --- the spin-orbit
(SO) interaction. Although this interaction is relativistic in
nature, it nevertheless results in a significant modification to
the band structure of Q1D systems (see, e.g.,
experimental~\cite{exper} and theoretical~\cite{theor} papers). In
strictly 1D and 2D systems, the role of the SO interaction
basically reduces to a ``horizontal'' (in momentum space) splitting
of the energy branches corresponding to spin-up and spin-down
electron states. In contrast to this, the existence of quantised
transverse energy subbands in quantum wires brings about an
additional important manifestation of the SO coupling: the absence
of a vertical symmetry axis for each spin branch (see
Fig.~\ref{fig1} and Ref.~\cite{MB}). This absence results in a
breakdown of chiral symmetry so that the electron Fermi
velocities become dependent on the direction of motion. This
effect monotonically increases as the SO coupling is enhanced. As
applied to interacting systems, the absence of chiral symmetry
raises two fundamental questions: how does the system respond to
the asymmetry of the single-particle spectrum; and what is the
fate of the spin-charge separation in the presence of the SO
coupling? It is the aim of this Letter, to establish a realistic
model for the SO coupling in interacting 1D systems which answers
these questions, and to provide experiment with a set of
quantitative predictions which will allow their verification.

In constructing a Hamiltonian for such a system, we consider the
case where the Fermi energy $E_F$ is sufficiently small, such that
only the lowest energy subband in a quantum wire is partly filled,
while all the others are empty. This regime has proven to be the
richest in non-trivial experimental results~\cite{tubes,anomaly},
and the SO effects are expected~\cite{MB} to be most pronounced
here. As a natural way of capturing the essential physics in a quantum wire, we
suggest the
 use of a modification to the Tomonaga-Luttinger model that takes
into account the asymmetric single-particle spectrum in
Fig.~\ref{fig1}. Namely, we take the Hamiltonian to have the form
$H = H_0 + H_{int}$, where $H_0$ describes the linearized spectrum,
\begin{eqnarray}
H_0 & = & -iv_1\int
dx\;\left(\psi^\dagger_{R,\uparrow}\partial_x\psi_{R,\uparrow}-
\psi^\dagger_{L,\downarrow}\partial_x\psi_{L,\downarrow}\right)
\nonumber \\ && - iv_2\int
dx\;\left(\psi^\dagger_{R,\downarrow}\partial_x\psi_{R,\downarrow}-
\psi^\dagger_{L,\uparrow}\partial_x\psi_{L,\uparrow}\right),
\label{H0}
\end{eqnarray}
and $H_{int}$ is responsible for the EE interaction:

\begin{eqnarray}
H_{int} &=& g_2\int dx\;\psi^\dagger_{R,\uparrow}\psi_{R,\uparrow}
\psi^\dagger_{L,\downarrow}\psi_{L,\downarrow} \nonumber \\
&+&g_4\int dx\;\psi^\dagger_{R,\uparrow}\psi_{R,\uparrow}
\psi^\dagger_{R,\downarrow}\psi_{R,\downarrow}+\left(R
\leftrightarrow L\right). \label{Hint}
\end{eqnarray}
The operators $\psi_{r,s}(x)$ ($r=R,L$; $s=\uparrow,\downarrow$)
annihilate spin-up ($\uparrow$) and spin-down ($\downarrow$)
 electrons near the
right ($R$) and left ($L$) Fermi
 points. In Eq.~(\ref{H0}) the electron
velocities
 $v_{1,2}=\partial E(p_{1,2})/\partial p$ are derived from the
dispersion law in Fig.~\ref{fig1}. In general, $v_1 \neq v_2$ as
long as the spin-orbit interaction is finite~\cite{MB}.

In the interaction Hamiltonian (\ref{Hint}) we include only
forward scattering contributions, while the backward and Umklapp
scattering are left out. Neglecting the Umklapp scattering is
known~\cite{Voit} to be safe if energy bands are far from being
half-filled, which is exactly the case, e.g., in quantum wires
patterned in semiconductor heterostructures~\cite{sp}. As regards
the backscattering, it can be shown that, similar to the case of
zero SO coupling~\cite{Voit}, it renormalizes to zero for
repulsive interactions and therefore can also be omitted. In this
letter, we concentrate on the major tendencies arising from the SO
coupling in interacting electron systems, therefore we assume
momentum-independent interaction potentials $g_2>0$ and $g_4>0$ in
Eq.~(\ref{Hint}), which corresponds to a well-screened repulsion
between electrons.

The Hamiltonian (\ref{H0}), (\ref{Hint}) is reminiscent of that
for the multi-component Tomonaga-Luttinger model~\cite{Penc,Voit}
which consists of mutually interacting Luttinger liquids with
different Fermi velocities. However, in contrast to our case, the
model~\cite{Penc,Voit} assumes that each liquid has a {\it
symmetric} single-electron spectrum (as occurs, for example, with
Zeeman splitting~\cite{Aoki}) and therefore, as we will show, it
describes qualitatively different physical behaviour.

We study the interacting model (\ref{H0}), (\ref{Hint}) with the
help of the bosonization technique~\cite{boson}. Within this
formalism, fermionic operators $\psi_{r,s}(x)$ are represented as
$\psi_{r,s}(x) \sim \exp(-i\Phi_{r,s})$, where the operators
$\Phi_{r,s}(x)$ are linear combinations of so-called phase fields
$\varphi_{\rho(\sigma)}(x)$ and $\Pi_{\rho(\sigma)}(x)$ that obey
the bosonic canonical commutation relations (see
Refs.~\cite{boson} for details). In terms of these phase fields,
the Hamiltonian (\ref{H0}), (\ref{Hint}) takes the bosonized form:
\begin{eqnarray}
H = \frac{1}{2\pi}\int dx & \Bigl\{ & v_\rho K_\rho
(\pi\Pi_\rho)^2+ \frac{v_\rho}{K_\rho}(\partial_x\varphi_\rho)^2
\nonumber \\ &+&v_\sigma K_\sigma (\pi\Pi_\sigma)^2+
\frac{v_\sigma}{K_\sigma}(\partial_x\varphi_\sigma)^2 \nonumber
\\ &+&\delta v\left[(\pi\Pi_\rho)\partial_x\varphi_\sigma+
(\pi\Pi_\sigma)\partial_x\varphi_\rho\right]\Bigr\}, \label{Hbos}
\end{eqnarray}
where
\begin{equation}
v_0=(v_1+v_2)/2, \qquad \delta v=v_2-v_1 \label{notation}
\end{equation}
and $v_{\rho(\sigma)}=\left[(v_0 \pm
g_4/2\pi)^2-(g_2/2\pi)^2\right]^{1/2}$ and
$K_{\rho(\sigma)}=\left[(2\pi v_0 \mp g_2 \pm g_4)/(2\pi v_0 \pm
g_2 \pm g_4)\right]^{1/2}$. The first (second) line in
Eq.~(\ref{Hbos}) describes free propagation of charge (spin)
density wave with the velocity $v_{\rho(\sigma)}$ and
``stiffness'' coefficient $K_{\rho(\sigma)}$. The third line in
Eq.~(\ref{Hbos}) is proportional to the velocity difference
$\delta v$ and therefore represents the strength of the SO
interaction.  In the Tomonaga-Luttinger model, where $v_1=v_2$,
this term is absent and this results in a decoupling of the charge
and spin degrees of freedom: the so-called {\it spin-charge
separation}~\cite{Voit}. This decoupling inhibits the existence of
quasi-particles with spin $1/2$ and charge $-e$, the basic
excitations of a Fermi liquid, and gives rise to a different state
of matter, the Luttinger liquid, which has bosonic excitations in
the form of independent spin and charge density waves.

As the SO interaction is switched on, i.e. at $\delta v \neq 0$,
the third term in Eq.~(\ref{Hbos}) starts to affect the dynamics
of the system. It couples together the $\rho$- and $\sigma$-fields
and thereby {\it destroys} this spin-charge separation. A similar
effect is also found in a spinful Luttinger model in a magnetic
field~\cite{Aoki}. In our case, the asymmetry of the
single-electron spectrum in Fig.~\ref{fig1} results in a different
mechanism for the violation of spin-charge separation from
Ref.~\cite{Aoki}. We stress however, that our model is exactly
solvable since the Hamiltonian (\ref{H0}), (\ref{Hint}) is
quadratic in the phase fields. Here we choose to solve it within
the functional integration formalism in imaginary (Matsubara)
time~\cite{funint}.

The most profound effects of the SO coupling on interacting Q1D
systems can be seen in the behaviour of single-particle
characteristics, such as the spectral function $\rho_r(q,\omega)$,
\begin{equation}
\rho_r(q,\omega)=-\frac{1}{\pi}\sum_s{\rm
Im}\; G_{r,s}^{(ret)}(q,\omega), \qquad r=R,\ L, \label{rho-def}
\end{equation}
and the density of states $N(\omega)=(1/2\pi)\sum_{r}\int
dq\;\rho_r(q,\omega)$. Here $q$ and $\omega$ are the momentum and
energy of elementary excitations and $G_{r,s}^{(ret)}(q,\omega)$
is the Fourier transform of the retarded Green function

\begin{eqnarray}
G_{r,s}^{(ret)}(x,t) &\equiv& ·
i\theta(t)\left<\left\{\psi_{r,s}(x,t),
\psi^\dagger_{r,s}(0,0)\right\}\right> \nonumber \\
&=&-i\theta(t)\left[G_{r,s}(x,t)+G_{r,s}(-x,-t)\right],
\label{G-def}
\end{eqnarray}
where $\theta(t)$ is the Heaviside step function and
$G_{r,s}(x,t)=\left.\left<\psi_{r,s}(x,\tau)
\psi^\dagger_{r,s}(0,0)\right>\right|_{\tau \to it} $ with $\tau$
being the Matsubara time. Leaving detailed calculations to a
regular article, we give here an exact expression for
$G_{r,s}(x,t)$ at $T=0$:
\end{multicols}
\widetext
\begin{equation}
G_{r,s}(x,t)=\frac{\exp(ip_{r,s}x)}{2\pi\Lambda}
\left(\frac{\Lambda}{\Lambda+i(u_1t-x)}\right)^{\theta_1^+}
\left(\frac{\Lambda}{\Lambda+i(u_1t+x)}\right)^{\theta_1^-}
\left(\frac{\Lambda}{\Lambda+i(u_2t-x)}\right)^{\theta_2^+}
\left(\frac{\Lambda}{\Lambda+i(u_2t+x)}\right)^{\theta_2^-}.
\label{G}
\end{equation}

\begin{multicols}{2}
\narrowtext A small (of the order of the lattice parameter) length
scale $\Lambda$ defines ultraviolet cut-off~\cite{boson}. From
Eq.~(\ref{G}) it follows that although the SO coupling destroys
the spin-charge separation, it nevertheless preserves the
anomalous scaling of correlation functions. At $\delta v=0$,
Eq.~(\ref{G}) coincides with the well-known
expression~\cite{Voit}. The exponents $\theta_i^\pm$ ($i=1,2$) are
determined by values of $v_{\rho(\sigma)}$, $K_{\rho(\sigma)}$,
$\delta v$ and depend on the chiral $r$ and spin $s$ indices. For
arbitrary interaction constants $g_2$ and $g_4$, the expressions
for $\theta_i^\pm$ are rather lengthy and here we give them for
the realistic case of equal interaction strength between all the
spectral branches, i.e. for $g_2=g_4 \equiv (\pi v_0)\beta$:
\begin{equation}
\theta_i^\pm(r,s) =
(-1)^i\xi_i^\pm(r,s)\eta_i/2(\eta_1^2-\eta_2^2), \label{theta}
\end{equation}
\begin{eqnarray}
\xi_i^\pm(r,s)&=&\pm
r\left[(1+rs\epsilon/2)^2/\eta_i-\eta_i\right] -(1-rs\epsilon/2)
\nonumber \\ &+&\left[(1+rs\epsilon/2)
(1-\epsilon^2/4)-\beta^2/2\right]/\eta_i^2. \nonumber
\end{eqnarray}
Here $\epsilon=\delta v/v_0$, and on the r.h.s. of the last
equation $r=\pm 1$ for right (left) spectrum branches and $s=\pm
1$ for spin-up (spin-down) electrons.

The quantities
 $\eta_{1,2} \equiv \eta_{1,2}(\epsilon)$ defined by
\begin{equation}
\eta_{1,2}^2 \equiv (u_{1,2}/v_0)^2 =1+\epsilon^2/4 \mp
\sqrt{\beta^2+\epsilon^2} \label{S12}
\end{equation}
are dimensionless velocities of the new {\it independent} bosonic
excitations that replace the spin and charge density waves,
respectively, as the single-electron spectrum becomes asymmetric
($\epsilon \neq 0$). Each of these new excitations are
superpositions of the old ones, they carry {\it both} charge and
spin, and have a sound-like spectrum $\omega=u_{1,2}q$. For
$\epsilon=0$, we have $u_{1,2}=v_{\sigma(\rho)}$ and return to the
spin-charge separation. Eq.~(\ref{S12}) demonstrates that
increasing $\epsilon$ at constant $\beta$ pushes $u_1$ and $u_2$
away from $v_\sigma$ and $v_\rho$ as well as from each other: one
of the excitations monotonically accelerates ($u_2$ grows) while
the other monotonically slows down ($u_1$ decreases). This effect
becomes more pronounced in systems with stronger EE interaction.
Since the parameter $\epsilon$ is controlled by the SO interaction
and therefore by the lateral electrical confinement, the strength
of this confinement must affect drastically the dynamics of
elementary excitations in quantum wires. Further increasing
$\epsilon$ eventually results in a vanishing of the velocity $u_1$
at the ``critical'' point $\epsilon_0=2\sqrt{1-\beta}$ and in a
possible phase transition accompanied by phase
separation~\cite{Voit}. This phenomenon is beyond the scope of
this Letter.

\begin{figure}
\begin{center}
\epsfig{file=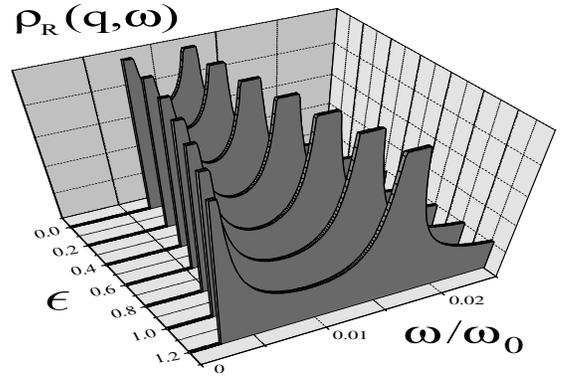,width=3.00in,height=2.00in}
\caption{Spectral function $\rho_R(q,\omega)$ (in arbitrary units)
for a fixed $q=0.01\Lambda^{-1}$ and several values of $\epsilon$
and for $\beta=0.5$ ($\omega_0 \equiv v_0/\Lambda$). Note that
$\rho_L(q,\omega)=\rho_R(q,-\omega)$.} \label{fig2}
\end{center}
\end{figure}

Eq.~(\ref{G}) allows one to calculate the spectral function
$\rho_r(q,\omega)$ which can be obtained in magnetotunneling
experiments~\cite{Kardynal,Altland}. Fig.~\ref{fig2} shows the
spectral function $\rho_R(q,\omega)$ for right-moving electrons in
the range of energies $\omega$ where all singularities of
$\rho_R(q,\omega)$ are located. At each value of
$\epsilon<\epsilon_0$, the spectral function has two singular
points corresponding to the bosonic excitations with velocities
$u_1$ and $u_2$ [see Eq.~(\ref{S12})]. In full accordance with
Eq.~(\ref{S12}), the distance between the singular points grows
with $\epsilon$. The singularities of $\rho_R(q,\omega)$ are
governed by power laws with exponents depending on
$v_{\rho(\sigma)}$, $K_{\rho(\sigma)}$, and $\epsilon$. The
dependence on $\epsilon$ is such that one of the singularities
becomes sharper while the other becomes smoother as $\epsilon$
grows. It is noteworthy that the dependence of $\rho_R$ on $q$ at
fixed $\omega$ is very similar to that shown in Fig.~\ref{fig2}
and exhibits the same tendencies as a function of $\epsilon$.

From Eqs.~(\ref{rho-def}), (\ref{G-def}), and (\ref{G}) one can
also calculate the single-particle density of states $N(\omega)$.
This quantity may be obtained in tunneling measurements since the
tunneling current, e.g., between a wide metal and an interacting
quantum wire is proportional to $N(eV)$, where $V$ is an applied
voltage (see Ref.~\cite{Matveev}). It can also be determined in
angle-integrated photoemission experiments~\cite{Dardel}. For
$\omega \to 0$ we obtain the following behaviour:
\begin{equation}
\frac{N(\omega)}{N_0} = \frac{1}{4}\sum_{r,s}
\frac{(\omega/\omega_0)^{\theta_1 + \theta_2 -1 }}
{\eta_1^{\theta_1} \eta_2^{\theta_2}\ \Gamma(\theta_1 +
\theta_2)}, \quad \theta_i=\theta_i^+  +\theta_i^-. \label{N-as}
\end{equation}
Here $\omega_0=v_0/\Lambda$ is the natural energy unit of the
order of the Fermi energy, $N_0=2/\pi v_0$ is the constant density
of states at $\beta=\epsilon=0$, and $\Gamma(x)$ is the Gamma
function.

Fig.~\ref{fig3}a demonstrates the effect of the EE interaction on
the density of states $N(\omega)$ at a fixed strength of the SO
coupling. At zero EE interaction ($\beta=0$), the function
$N(\omega)$ is a constant, $N(\omega)/N_0=1/(1-\epsilon^2/4)$. As
the EE interaction is switched on ($\beta \neq 0$), it starts to
affect the formation of the lowest lying excitations and
$N(\omega)$ takes on a power-law behaviour in the vicinity of
$\omega=0$. The width of this power-law interval becomes
progressively larger as $\beta$ grows and leads to a monotonic
suppression of the density of states. This effect is also present
for zero SO coupling~\cite{Voit}.
\begin{figure}
\begin{center}
\epsfig{file=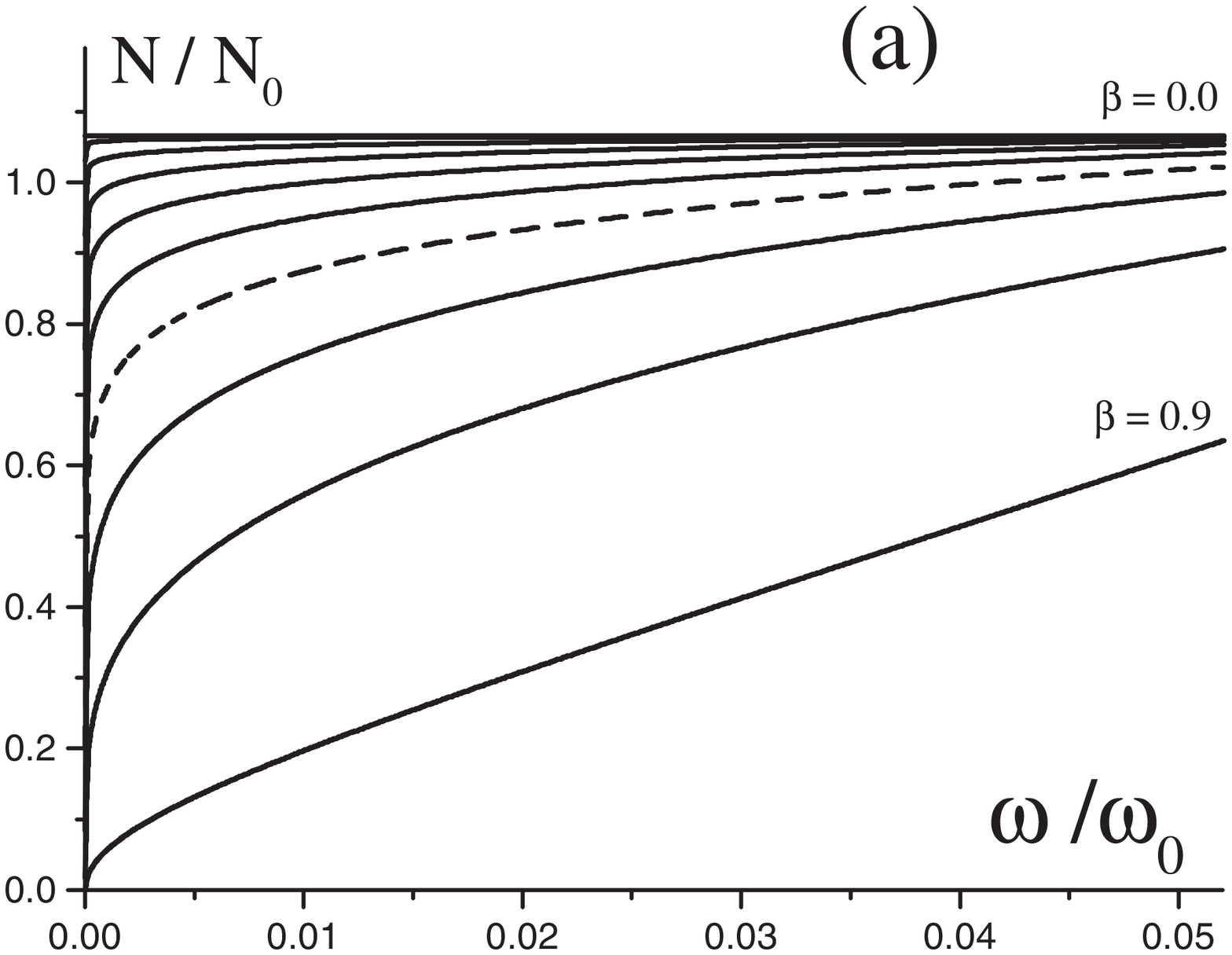,width=3.00in,height=2.00in}
\end{center}
\end{figure}
\vspace{-6ex}
\begin{figure}
\begin{center}
\epsfig{file=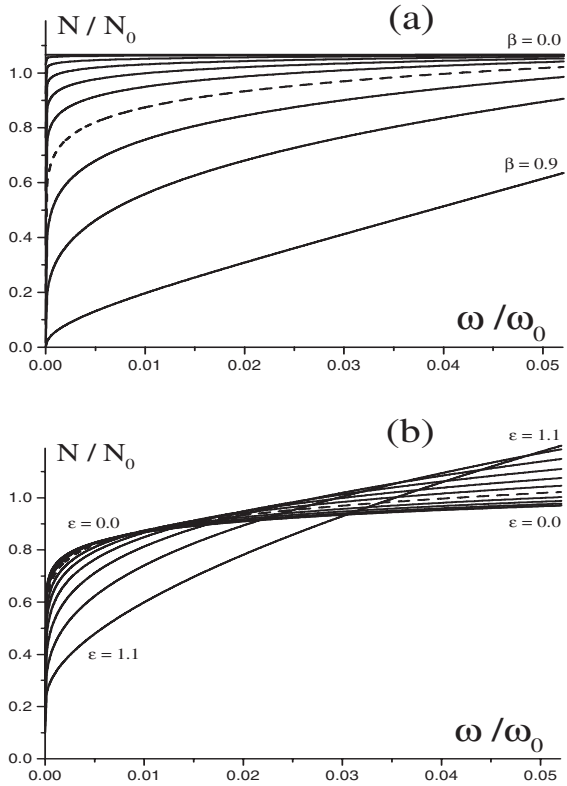,width=3.00in,height=2.00in}
\caption{Normalised single-electron density of states: (a)
$\epsilon=0.5$ and $\beta=0.0, 0.1,\ldots ,0.9$; (b) $\beta=0.6$
and $\epsilon=0.0,0.1,\ldots ,1.1$. The dashed curves correspond
to $\epsilon=0.5$, $\beta=0.6$.} \label{fig3}
\end{center}
\end{figure}

Fig.~\ref{fig3}b shows the evolution of the function $N(\omega)$
for fixed $\beta$ as the SO coupling is varied. For large values
of $\omega$, where the role of EE interactions is not significant,
the magnitude of $N(\omega)$ increases as $\epsilon$ grows.
However, at small values of $\omega$, where the nature of the
elementary excitations is essentially dictated by EE interactions,
the effect of the SO coupling on $N(\omega)$ is qualitatively the
same as that of the EE interaction, that is, increasing $\epsilon$
leads to a suppression of the density of states. We emphasize that
in the standard multi-component Tomonaga-Luttinger
model~\cite{Penc,Voit} as well as in spin-polarized Luttinger
liquids~\cite{Aoki}, there is no interval of $\omega$ where the
density of states is suppressed by increasing the velocity
difference between spectral branches. The existence of such an
interval is a unique manifestation of the SO coupling in quantum
wires.

The characteristic dependence of the spectral function
$\rho(q,\omega)$ and the density of states $N(\omega)$ on the
microscopic parameter $\epsilon$, for fixed $\beta$, should enable
one to extract both quantities from experiment and thus determine
how strong the EE and SO interactions are. Such dependencies can
be obtained experimentally by changing the SO coupling directly by
varying an electric field perpendicular to a quantum wire
(quantum-well field) as it was done, e.g., in Ref.~\cite{Nitta}.

Theoretical calculations \cite{MB} of the electron bandstructure
modified by the SO coupling indicate that the value of $\epsilon$
in typical Q1D semiconductor systems should be $\sim 0.1$ -- $0.2$
and appears sufficiently large to observe the principal tendencies
in the behaviour of $\rho(q,\omega)$ and $N(\omega)$ caused by the
SO coupling.

In conclusion, we have formulated and solved analytically the
problem of the interplay between EE and SO interactions in Q1D
electron systems and found that: (i) the spin-charge separation
breaks down while the anomalous scaling of correlation functions
is preserved; (ii) the single-particle characteristics, such as
the spectral function and the density of states, are essentially
modified and controlled by the strength of the SO coupling; (iii)
varying the SO coupling with the external electric field can be
used to extract the microscopic parameters of quantum wires.

AVM thanks the ORS, COT, and Corpus Christi College for financial
support. KVS and CHWB thank the EPSRC for financial support.


\end{multicols}
\end{document}